\documentclass[a4paper,article,fleqn,usenatbib]{mnras}

\usepackage{newtxtext,newtxmath}
\usepackage[T1]{fontenc}
\usepackage{ae,aecompl}

\usepackage{graphicx}	
\usepackage{amsmath}	
\usepackage{amssymb}	
\usepackage{float}
\usepackage{tabularx} 
\usepackage{ctable}
\usepackage{multicol} 
\usepackage{geometry}

\title[Asymmetric Ejecta of Cool Super/Hypergiants in Wd1]{Asymmetric Ejecta of Cool Supergiants and Hypergiants in the Massive Cluster Westerlund 1}

\author[Andrews, H et al.]{Andrews, H.,$^{1}$\thanks{E-mail: holly.andrews.16@ucl.ac.uk (UCL)} Fenech, D.,$^{1}$ Prinja, R. K.,$^{1}$, Clark, J. S.,${^2},$ and  Hindson, L.${^3}$ 
\\
$^{1}$University College London, Gower St, Bloomsbury, London WC1E 6BT \\
$^{2}$Open University, Walton Hall, Kents Hill, Milton Keynes MK7 6AA \\
$^{3}$University of Hertfordshire, Hatfield, Hertfordshire, AL10 9AB }
\date{Accepted 2018 March 17. Received 2018 February 26; in original form 2018 January 17}

\pubyear{2018}

\begin{document}
\label{firstpage}
\pagerange{\pageref{firstpage}--\pageref{lastpage}}
\maketitle

\begin{abstract}
We report new 5.5$\,$GHz radio observations of the massive star cluster Westerlund 1, taken by the Australia Telescope Compact Array, detecting nine of the ten yellow hypergiants (YHGs) and red supergiants (RSGs) within the cluster. Eight of nine sources are spatially resolved. The nebulae associated with the YHGs Wd1-4a, -12a and -265 demonstrate a cometary morphology - the first time this phenomenon has been observed for such stars. This structure is also echoed in the ejecta of the RSGs Wd1-20 and -26; in each case the cometary tails are directed away from the cluster core. The nebular emission around the RSG Wd1-237 is less collimated than these systems but once again appears more prominent in the hemisphere facing the cluster. Considered as a whole, the nebular morphologies provide compelling evidence for sculpting via a physical agent associated with Westerlund 1, such as a cluster wind.
\end{abstract}




\section{Introduction}
\defcitealias{Dougherty2010}{D10}
\defcitealias{Fenech2018}{Fenech, submitted}

Mass-loss is a fundamental agent in the evolution of massive stars, stripping away their H-rich outer layers as they evolve away from the main sequence. Stars in the $\sim$$\,$10$\,$-$\,$40$\,$$M_{\odot}$ mass-range encounter a cool red supergiant (RSG) and/or yellow hypergiant (YHG) phase, expected to be associated with significant mass-loss driven by either steady-state, cool, dense stellar winds, or impulsive ejections of material$\,$\citep[cf.][]{Shenoy2016}. Unfortunately our understanding of the winds from cool supergiants and hypergiants is not yet deeply developed and uncertainties remain in commonly used prescriptions adopted by evolutionary codes \citep[e.g.][]{vanLoon2005}. Indeed, the situation becomes progressively worse as more massive examples are considered, since their increased rarity decreases the empirical constraints available. Nevertheless the handful of known examples suggest `quiescent' mass-loss rates of $\geq10^{-5}\,M_{\odot}$, which may increase by an order of magnitude or more in eruptive events$\,$\citep[][\citealt{Shenoy2016}]{Smith2014}.

In many cases, mass-loss rates are inferred from the properties of the dust rich ejecta associated with a number of RSGs and YHGs. However, the central stars are too cool to ionise either their winds or nebulae and so ejecta masses are inferred from dust mass; a methodology susceptible to uncertainties in the dust-to-gas ratio. This is unfortunate since the circumstellar matter resulting from winds and ejections provides an excellent probe of the mass-loss history of the star. Moreover, the nature of the circumstellar environment around these stars directly affects the observational properties of the eventual SNe via interaction with the SN blast-wave. It is expected that stars with initial masses between $\sim$9$\,$-$\,$18 M$_{\odot}$ evolve into RSGs that are direct progenitors for SN IIP explosions, while more massive  YHGs yield SN IIL/b \citep[e.g.][]{Groh2013}; if one is to understand the nature of these (most) common core-collapse events one must first understand the environment they occur within.

In this letter we highlight results from high-sensitivity ATCA radio observations of the stellar ejecta around four RSGs and five YHGs in the massive star cluster, Westerlund 1 (Wd1). These results have been compared in a qualitative manner to 100$\,$GHz ALMA observations taken in 2015. Westerlund 1 is the most massive starburst cluster in the Milky Way and with an age of $\sim5$$\,$Myr it contains an unprecedentedly rich population of very massive stars \citep[$>200$ with masses $\geq30\,M_{\odot}$;][]{Clark2005}. Such a cohort yields a strong cluster UV radiation field, which results in the ionisation of otherwise neutral ejecta around the cluster RSGs and YHGs; allowing a {\em direct} determination of nebular masses. Building on the results of \citep[][henceforth \citetalias{Dougherty2010}]{Dougherty2010}, we focus on the detailed morphology of the RSG and YHG ejecta, which provide an outstanding opportunity to examine evidence for asymmetries that may  betray the action of a cluster wind.

\begin{table*}
	\vspace{-10pt}
	\centering
	\caption{Listed here in order: source name, spectral type, flux densities for the 750$\,$m and 6$\,$km pointing from the 5.5$\,$GHz measurements (in$\,$mJy), and dimensions of the ejecta (in$\,$arcseconds), with the major axis (top) and the minor axis (bottom). Also shown is the source size (in parsecs, with an assumed distance of 5$\,$kpc), the volume of the ejecta (in cm$^{3}$), and the ejecta mass estimates (in x$\,$10$^{-3}$ M$_{\odot}$).}
	\label{tab:table1}
	\begin{tabularx} 
		{\linewidth} 
		{cccccccc}
		\hline
		Star & Spectral & Flux Density	& Flux Density & Ejecta   &  Source Size   & Volume  				 &  Ejecta Mass     \\
		&  Type    	    	& 750m (mJy) & 6km (mJy)) & Dimensions ($\arcsec$)	& Deconvolved (pc)								  &($\times 10^{53}$cm$^{3}$) 	&         ($\times$10$^{-3}$M$_{\odot}$)           	\\
		\hline
		
		W4a	 & F3 Ia+ 	& 9.09 $\pm$ 1.50	& 3.61 $\pm$ 0.04	  & 10.7 $\pm$ 3.2 &0.26 $\pm$ 0.08  & 6.72 $\pm$ 5.63		& 97.7 $\pm$ 41.0    \\
		        &              &                              &                               & 6.0 $\pm$ 4.7                       &0.15 $\pm$ 0.11    &                               &   \\ 
		
		W12a & F1 Ia+ & 6.99 $\pm$ 0.67		&2.84 $\pm$ 0.05  & 7.5 $\pm$ 2.2  & 0.18 $\pm$ 0.05 	& 4.55 $\pm$ 2.67								 & 70.6  $\pm$ 20.7  \\
		          & 	    &	                           &					            & 5.9$\pm$ 3.0	                        & 0.14 $\pm$ 0.07	                           &								           &  \\

		W265 & F5 Ia+  & 2.99 $\pm$ 0.53	 &1.78 $\pm$ 0.04 & 11.1 $\pm$ 4.4 & 0.27 $\pm$ 0.11 &	1.62 $\pm$ 1.49 		& 27.6 $\pm$ 12.7       		\\
		           &            &                               &                           & 2.9 $\pm$ 2.4 &   0.02 $\pm$ 0.06           &                                 &  								 \\
		           
		W16a		& A2 Ia+ &	-- & 3.62 $\pm$ 0.34	  & 3.69 $\pm$ 0.55 & 0.089 $\pm$ 0.013  & 0.425 $\pm$ 0.105	& 15.5 $\pm$ 1.9   \\
		           & &          &                               & 2.57 $\pm$ 0.51                       &0.062 $\pm$ 0.012    &                               &   \\ 
		           
	   W32  	& F5 Ia+ & 	-- & 0.367 $\pm$ 0.052  & -- & --	& 0.121							 & $\sim$ 2.63 \\
		           &&   &					            &  --                &  --                &								           &  \\
		
		W20 & M5 Ia & 30.6 $\pm$ 6.7	&5.30 $\pm$ 0.30 & 18.3 $\pm$ 5.0 & 0.44 $\pm$ 0.12& 53.9 $\pm$ 21.3		& 508 $\pm$ 100		\\
		         &          &             	               &	                      &	13.0 $\pm$ 3.7 & 0.32 $\pm$ 0.10	&  & \\
		
		W26 & M5 Ia  & 221 $\pm$ 19.2 	 &142 $\pm$ 26	& 10.8 $\pm$ 2.2  & 0.26 $\pm$ 0.05	 & 4.71 $\pm$ 2.20 &  403 $\pm$ 94	 \\
		
		&			&					& & 5.0 $\pm$ 2.1 & 0.12 $\pm$ 0.05 & &   \\
		
		W237 & M3 Ia  & 9.83 $\pm$ 0.47	 & 2.51 $\pm$ 0.05	& 8.2 $\pm$ 0.5	  & 0.199 $\pm$ 0.012 & 3.57 $\pm$ 1.38	& 74.1 $\pm$ 14.3					\\
		
		&		&	&			&		5.0$\pm$ 1.9	 	& 0.12 $\pm$ 0.05  &	&   \\
	    W75 	& M5 Ia  &  --   & 0.530 $\pm$ 0.056 & 1.86 $\pm$  0.53 & 0.045 $\pm$ 0.012 &	0.065 $\pm$ 0.028		& 2.34 $\pm$ 0.49    		\\
	                                   &&     &                           & 1.42 $\pm$ 0.44 &  0.034 $\pm$ 0.010         &                                 &  								 \\

		\hline
		\label{table:1} 
		
	\end{tabularx}
\end{table*}
\section{Observations}
Observations were taken of Westerlund 1 using ATCA (Australia Telescope Compact Array). ATCA consists of 6 antennas with diameters of 22$\,$m. The observations took place from October to December 2015, in three different configurations each defined by their widest baseline. The 6$\,$km configuration was observed over 27th - 29th October, the 1.5km  configuration over 25th - 27th November and the 750$\,$m configuration over 14th - 15th December. Follow up observations on the 3rd June 2016  with the 1.5$\,$km configuration were carried out, due to issues caused by poor weather during the original observing period.
 
Data was collected over two spectral windows, centered at 5.5 and 9$\,$GHz, each with a bandwidth of 2$\,$GHz and 2048 1 MHz-wide channels. The total on-source integration time per pointing was $\sim$16 hours. The data contained observations of a bandpass calibrator, 1934-638, a phase calibrator, J1636-4101 for the 6$\,$km configuration, or 1600-48 elsewhere, and a second calibrator 0823-500.
  
The data were flagged and calibrated with the use of {\sc{miriad}}$\,$\citep[][]{Miriad}. {\sc{miriad}} was also used to perform phase and target self calibration as well as final imaging of each configuration. The images were then primary beam corrected and exported to {\sc{casa}}$\,$(Common Astronomy Software Applications) for processing. The 6$\,$km and 750$\,$m images were analysed within {\sc{casa}} to determine flux densities and ejecta dimensions. The final set of images created had varying resolutions. The 6$\,$km pointing at 5.5$\,$GHz had a beam size of 2.92$\,$$\times$$\,$1.54$\arcsec$ (PA -2.50$^{\circ}$) and at 9$\,$GHz, a beam size of 1.71$\,$$\times$$\,$0.99$\arcsec$ (PA 0.79$^{\circ}$). Images created from the 750$\,$m pointing had a beam size of 13.79$\,$$\times$$\,$6.98$\arcsec$ (PA 5.04$^{\circ}$) at 5.5$\,$GHz, and 6.98$\,$$\times$$\,$4.44$\arcsec$ (PA 6.28$^{\circ}$) at 9$\,$GHz.

Calibration and imaging is still on-going for the 1.5$\,$km pointings at both spectral windows. A full combination of all the data is underway, with results from the concatenated dataset to be provided in a future work (Andrews, in prep).

\begin{figure}
	\vspace{-10pt}
	\centering \includegraphics[width=0.97\textwidth]{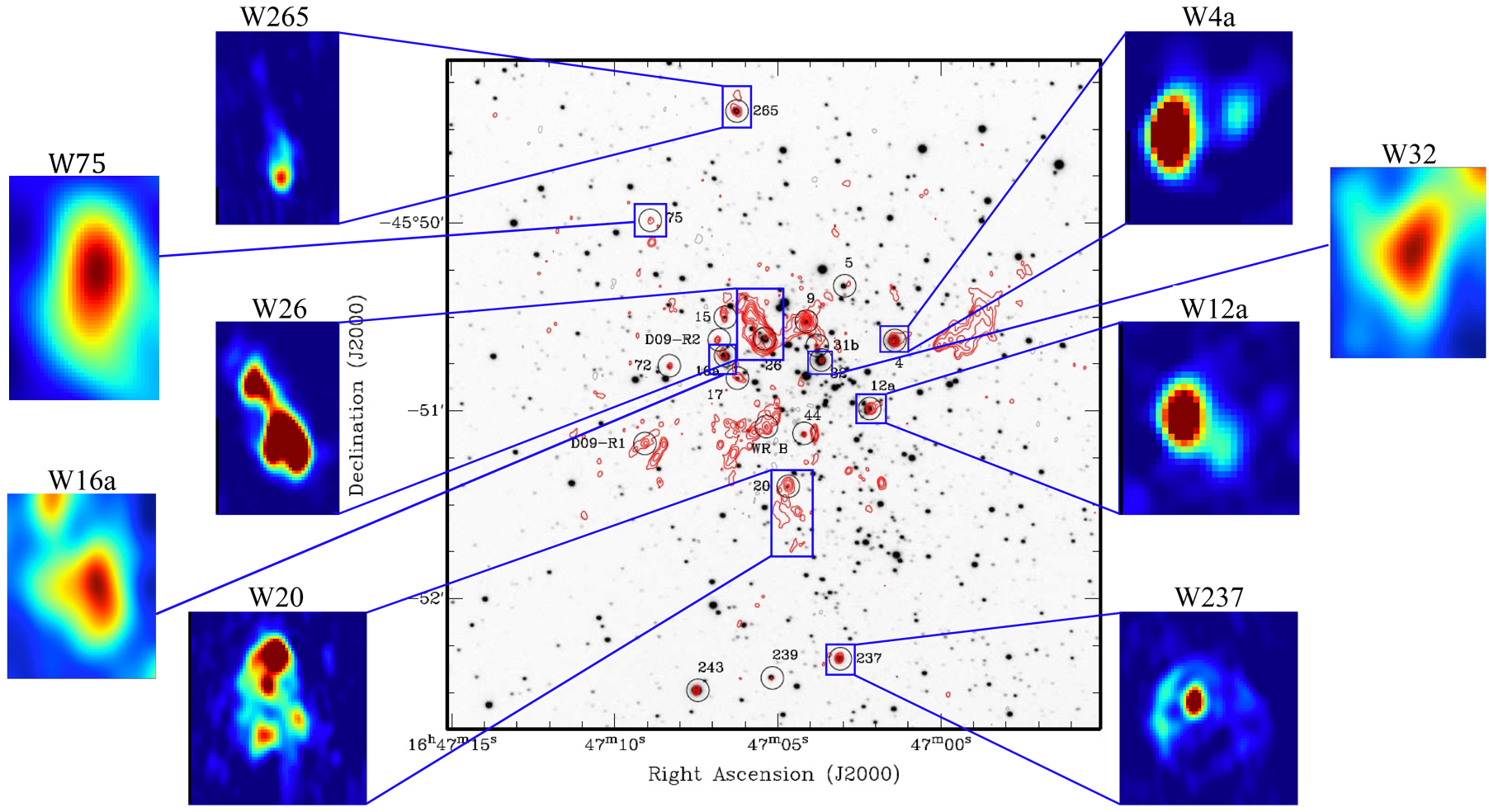}
	\caption{The central image shows the 2009 set of ATCA radio observations of Westerlund 1 \protect\citepalias[][]{Dougherty2010}. $\textit{Left}$ are Wd1-75, -16a, -265, -26 and -20, and $\textit{right}$ are Wd1-4, -12a, -237, and -32, with the blue lines and central boxes indicating their place within the cluster. This image demonstrates the clear link between the direction of the asymmetry and the placement of the stellar objects in the cluster.}
	\label{fig:2}
\end{figure}

In this letter we focus on the presence of asymmetric spatially extended emission associated with all but one of the cool super/hypergiants, seen in the 5.5$\,$GHz observations at the 6$\,$km pointing.$\,$\cite{Clark1998} previously identified a cometary nebula around the RSG Wd1-26, with \citetalias{Dougherty2010} confirming similar nebulosity around a second RSG Wd1-20, and an elliptical structure around a third RSG, Wd1-237, as well as more compact but still extended emission around three of the YHGs, Wd1-4a, -12a and -265. The greater sensitivity of our new observations reveals striking additional detail for these structures, while resolving the nebulae associated with the last RSG, Wd1-75 and two further YHGS, Wd1-16 and -32. This leaves only a single cluster YHG, Wd1-8, with no detection at any mm or radio frequency. In Fig \ref{fig:1} we present 5.5$\,$GHz maps of these six stars, while Fig \ref{fig:2} presents an optical/radio montage indicating the nebulae location and orientation.
\section{Properties of the Circumstellar Envelopes}
We present the bulk properties of the resolved nebulae in Table \ref{table:1}. The gaussian-fitting procedure {\sc{imfit}} in {\sc{casa}} was used to calculate the flux densities associated with the extended regions for each source. First, fits were performed on the 6km pointing at 9$\,$GHz (with the highest resolution) to determine the dimensions of the central sources. The total flux density present in the extended emission regions for each source was then found using data from the (lower resolution) 750$\,$m pointings at 5.5$\,$GHz. The final flux density was calculated by taking the total flux density and subtracting the flux present in the region covering the central compact source (as determined by the compact fit at 9$\,$GHz). For Wd1-26, the source fit would not converge for the central source dimensions, so the value given is the total flux density from running {\sc{imfit}} on the 750$\,$m 5.5$\,$GHz image. 

Ejecta dimensions were also taken from {\sc{imfit}}. The major and minor axes listed are the deconvolved (i.e. having removed the effect of the convolving beam) FWHM of the gaussian fit taken from the 750$\,$m 5.5$\,$GHz observations. Given the diffuse cluster emission and that of nearby sources, there is some difficulty in providing an accurate fit that includes as much of the source emission as possible but avoids incorporating emission from other stars. In comparison to the apparent emission (in Fig. $\ref{fig:1}$ and $\ref{fig:2}$), it is possible that the dimensions for Wd1-26 and Wd1-237 have been underestimated. Likewise, the dimensions determined for Wd1-20 may be marginally overestimated.

The emission associated with Wd1-16a, -32 and -75 is smaller in both size and brightness. It was therefore not possible to isolate these sources for fitting in the lower resolution image. Flux densities and nebula dimensions were found using source fits on the 6$\,$km 5.5$\,$GHz data instead. Wd1-32 was detected, but the source was only partially resolved, so the convolving beam size was taken to represent the ejecta dimensions.
 
The flux density values from the 750$\,$m pointing at 5.5$\,$GHz were used to calculate the ejecta mass. The mass value calculated is the ionised mass, ignoring the presence of neutral material. To estimate this mass, the emission measure, $E_{V}$, was calculated by rearranging the equation for an optically thin thermal plasma at an effective temperature $T$,
\setlength{\abovedisplayskip}{-5pt}
\begin{equation}
 S_{\nu} = 5.7 \times 10^{-56}\,T^{\frac{1}{2}}\,g\,D^{2}\,E_{V} \quad   \textrm{mJy} \label{eq:flux}
\end{equation}
where $D$ is the distance, set to 5$\,$kpc, and $g$ is the Gaunt factor, calculated from the frequency, temperature and the stellar metallicity$\,$\citep[][]{LeithererRobert1991}. Following \citetalias{Dougherty2010}, we adopt a temperature of 10$^{3}$$\,$K. 

$E_{V}$ is equal to the integrand of the electron number density,$\,$$n_{e}$$\,$, over the ejecta volume. The full geometry of the ejecta could not be directly determined, so assumptions were required to calculate the volume. As the inclination of the envelope is unknown, it was assumed that the line-of-sight was face-on. The ejecta envelope was assumed to be a uniformly filled ellipsoid, using the major and minor axes found in {\sc{imfit}}. The third axis could not be directly determined, so the ellipsoid was taken to be symmetrical around the major axis, using the minor axis as a representative value for the third dimension. The electron density and emission measure were then used to calculate the ionised mass, 
\setlength{\abovedisplayskip}{3pt}
\setlength{\belowdisplayskip}{3pt}
 \begin{equation}
 M_{E} = \mu m_{H} \left(\frac{E_{V}}{n_{e}}\right)
 \end{equation}
where $m_{H}$ is the atomic mass of hydrogen and $\mu$ is the mean atomic weight, taken to be 1.5 \citep[][]{LeithererRobert1995}. Errors for the ionised mass were found by propagating errors on the ejecta dimensions and flux densities, given in {\sc{imfit}}.

\section{Discussion}
In this study, high sensitivity 5.5$\,$GHz ATCA observations  were exploited to demonstrate the presence of asymmetric ejecta around cool RSGs and YHGs in the young massive cluster Westerlund 1. Very few ejection nebulae are known to be associated with highly luminous RSGs and still fewer with YHGs \citep[][]{Shenoy2016}; as a consequence the detection of nine surrounding some of the most extreme examples within the Galaxy \citep[initial mass $\sim40\,M_{\odot}$:][\citealt{Clark2014}]{Ritchie2010} provides a unique opportunity to study the dynamic mass-loss processes of these evolutionary phases.

\begin{figure}
	\hspace*{-0.3in}
	\vspace{-5pt}
  \includegraphics[width=1.07\textwidth]{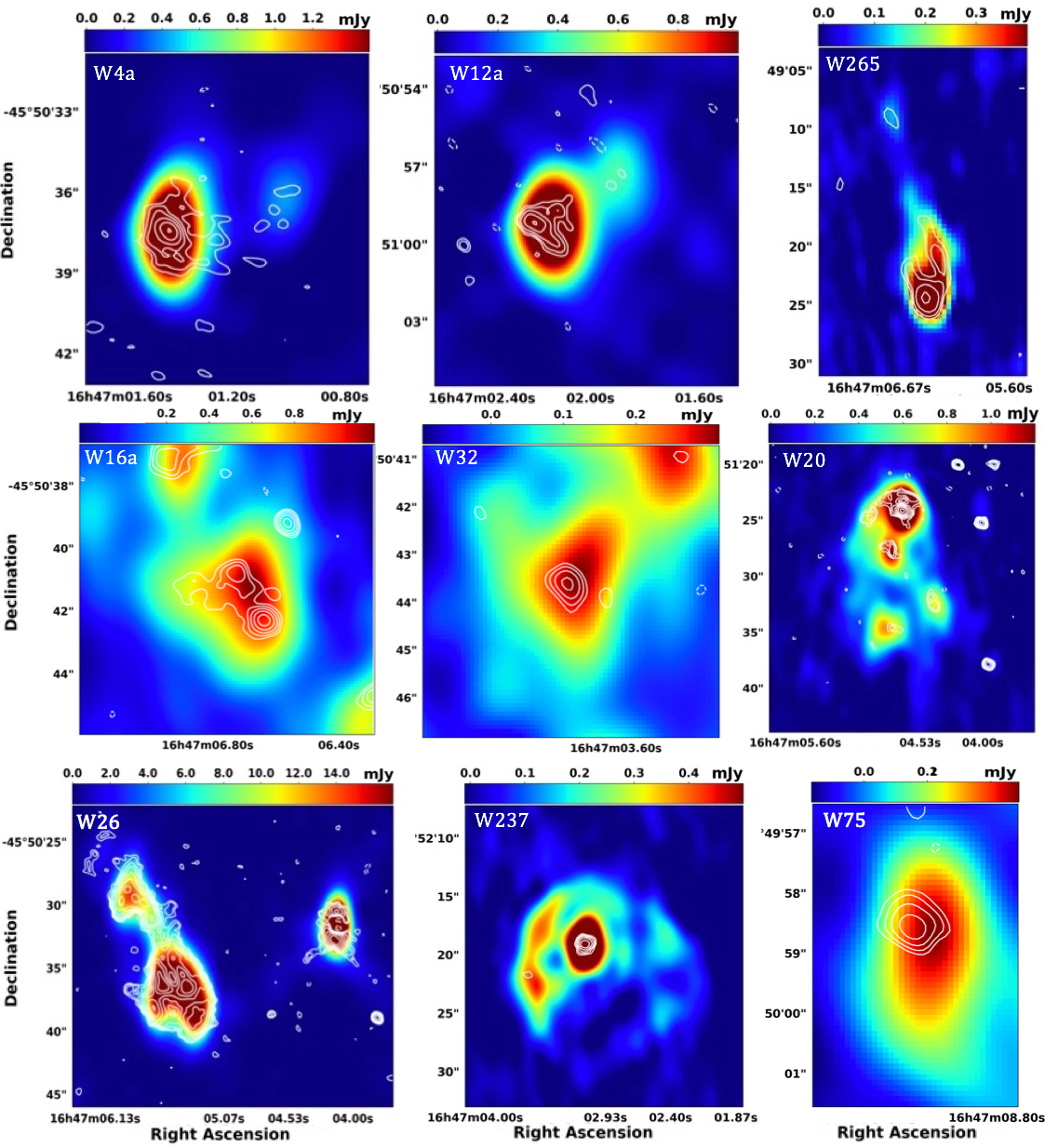}
  
  \caption{Shown are images of the YHGs and RSGs and their associated extended emission, from the 5.5GHz observations at 6km.  Sources are: the YHGs Wd1-4a $\textit{(top-left)}$, Wd1-12a $\textit{(top-centre)}$, Wd1-265 $\textit{(top-right)}$, Wd1-16a $\textit{(middle-left)}$ and Wd1-32 $\textit{(centre)}$. These images are followed by the RSGs, Wd1-20 $\textit{(middle-right)}$, Wd1-26 $\textit{(bottom-left)}$ and Wd1-237 $\textit{(bottom-centre)}$ and Wd1-75 $\textit{(bottom-right)}$. The colourscale shows the hypergiants from the 5.5GHz observations at the 6km pointing, with the overlaid contours from 100$\,$GHz ALMA observations of the cluster with a resolution of 750$\,$$\times$$\,$570$\,$mas \citepalias[][]{Fenech2018}. W265 was out of the field of view for ALMA, so is overlaid with contours of the 9$\,$GHz ATCA observations from the 6km pointing. The contour levels of the ALMA observations are -1, 1, 1.414, 2, 2.828, 4, 5.656, 8, 11.31, 16, 22.62, 32, 45.25, 64 and 90.50 $\times$$\,$3$\sigma$ ($\sigma$ = 52.4$\,$$\mu$Jy for Wd1-4a, 40.0$\,$$\mu$Jy for Wd1-12a, 24.3$\,$$\mu$Jy for Wd1-16a, 28.1$\,$$\mu$Jy for Wd1-32, 30.5$\,$$\mu$Jy for Wd1-20, 63.7$\,$$\mu$Jy for Wd1-26, 31.9$\,$$\mu$Jy for Wd1-237 and 26.2$\,$$\mu$Jy for Wd1-75). The 6km pointing at 9$\,$GHz has contour levels of 3, 6, 9, 12, 24, 48, and 192 $\times$ 15.0$\,$$\mu$Jy.} 

  \label{fig:1}
\end{figure}

The most arresting results from this study are the morphologies of the nebulae. Specifically, the cometary nebulae of Wd1-4a, 12a and 265 are the {\em only} known examples associated with YHGs. In the first two cases, the radio continuum emission is of greater physical extent than the mm-continuum emission, which is more reminiscent of a `bow-shock' morphology (Fig \ref{fig:1}); Wd1-265 was not covered by  the 100$\,$GHz ALMA observations but demonstrates the clearest cometary configuration of the three objects \citepalias{Fenech2018}. Bow-shocks have been associated with three (field) RSGs: Betelgeuse \citep[][]{Decin2012}, $\mu$ Cep \citep[][]{Cox2012}, and IRC-10414 \citep[][]{Gvaramadze2014}, with only a single example of a cometary nebula, around GC IRS7 \citep[][]{YusefZadeh1991}. Note that they are also common around AGB stars \citep{Cox2012}. In terms of physical extent and morphology, the nebulae of Wd1-26 is directly comparable to GC IRS7. Strikingly, the synthesis  of ALMA+ATCA data for Wd1-20 reveals that it shows both morphologies, with a compact nebular component positionally coincident with the star, clearly displaced from a bright arc of emission (the `bow-shock'), both of which are embedded in the cometary nebula. The ejecta associated with Wd1-237 appears to be a less collimated version of Wd1-20, with a similarly compact central source displaced from the geometrical centre of a quasi-spherical nebula with limb-brightening reminiscent of a bow-shock on the hemisphere orientated towards the cluster core. 

With the exception of Wd1-75, the nebulae associated with the RSGs appear systematically more massive than those of the YHGs (Table \ref{table:1}), with the YHG nebulae in turn appearing less massive than those of field YHGs, such as IRC +10420, thought to be in a post-RSG phase $\,$\citepalias[][]{Fenech2018}. There are two distinct scenarios for this difference in YHG and RSG nebular properties in Wd1: (i) the YHGs in this cluster are pre-RSG objects and not yet subject to the major mass loss of the RSGs, or (ii) the YHGs are post-RSG objects and hence the nebulae ejected in the RSG phase have been exposed to the cluster wind for longer and hence have suffered more erosion. It is tempting to attribute the differences between both classes of object within Wd1 as an evolutionary effect, with the YHGs yet to encounter the RSG phase and associated heavy, impulsive mass-loss. There is a general expectation of a systematic difference between wind properties of RSGs and YHGs, with those of the former thought to be slower, denser and cooler than the latter$\,$\citepalias[\citetalias{Dougherty2010},][]{Fenech2018}. If the difference in nebular properties results from the differing evolutionary phases, one might speculate that the lack of nebulosity associated with Wd1-8 is a result of the star only recently entering the YHG phase, although more prosaic explanations - i.e. the star is at a large displacement from the cluster core along our line of sight, such that the cluster radiation field is insufficient to ionise any ejecta - are possible. 

Nevertheless, the striking cometary nebulae of the RSGs Wd1-20 and 26, the `bow-shock' YHG nebulae and crucially the orientation of all these with respect to the cluster core compellingly argue for interaction with a physical agent(s) associated with the cluster. By analogy to classical runaways, motion through an ambient intracluster medium \citep[revealed via X-ray observations;][]{Muno2006} might sculpt the nebulosity, but this would requre {\em all} such objects to be infalling at supersonic velocities (`run-towards') which would appear contrived. 
\onecolumn 
\twocolumn
\onecolumn
\begin{multicols}{2}

It would seem more likely that their morphologies are due to interactions with an outflowing cluster wind, or the winds of ionising radiation fields of nearby massive stars, or supernova. The presence of a magnetar within Wd1 indicates at least one such recent event, with the mass of Wd1 implying a rate of occurence of 7000$\,$-$\,$13,000$\,$yr$^{-1}$$\,$\citep[][]{Muno2006}. \cite{Mackey2015} favour stellar wind interaction as the sculpting agent for the Wd1-26 nebula, but this analysis also demonstrates that models tailored for the immediate cluster environment of individual stars will be required to fully understand these systems. 	

Red supergiants and yellow hypergiants mark pivotal late phases in the evolution of massive stars. Furthermore, the properties of circumstellar matter associated with the progenitor RSGs and YHGs will affect the intra-cluster medium in Wd1, including the extent of cavities and filaments, which is pertinent to the morphology and development of the subsequent young supernova remnant \citep[e.g.][]{Vink2012}. Also, our results provide new perspectives on the manner by which these stars are directly affected by the massive cluster they are embedded in. 

A more complete analysis of the ATCA dataset, containing all the detected cluster members, is underway (Andrews, in prep), with direct quantitative comparisons between this data and the previous ATCA observations \citepalias[][]{Dougherty2010}, and recent ALMA observations \citepalias[][]{Fenech2018} to be made.

\section*{Acknowledgements}

H. Andrews wishes to acknowledge STFC for the funding of a PhD Studentship. D. Fenech wishes to acknowledge funding from a STFC consolidated grant (ST/M001334/1).

\bibliographystyle{mnras}
\bibliography{ref.bib} 








\bsp	
\label{lastpage}
\end{multicols}
\end{document}